\documentclass{ws-mplaMODIFIED}

\begin{document}

\markboth{Vasilis Niarchos}
{Phases of Higher Dimensional Black Holes}

\catchline{}{}{}{}{}

\title{\bf PHASES OF HIGHER DIMENSIONAL BLACK HOLES}

\author{\footnotesize VASILIS NIARCHOS}

\address{\hfill\\
Centre de Physique Th\'eorique, \'Ecole Polytechnique\\
Palaiseau, 91128, France\\
niarchos@cpht.polytechnique.fr}

\maketitle

\pub{}{\bf Abstract}

\begin{abstract}
We review some of the most striking properties of the phase diagrams of
higher dimensional black holes in pure gravity. We focus on static black 
hole solutions with Kaluza-Klein asymptotics and stationary black hole 
solutions in flat Minkowski space. Both cases exhibit a rich pattern of 
interconnected phases and merger points with topology changing transitions. 
In the first case, the phase diagram includes uniform and non-uniform black 
strings, localized black holes and sequences of Kaluza-Klein bubbles. In 
the latter case, it includes Myers-Perry black holes, black rings, black 
saturns and pinched black holes. 

\end{abstract}

\vspace{.2cm}
\centerline{\it Invited Review for Modern Physics Letters A}


\vspace{.1cm}
\section{Introduction}	

Our purpose in this short review is to summarize some of the recent progress
in the study of black hole solutions in higher dimensional classical general 
relativity. There are already several excellent reviews of complementary aspects
of this constantly growing subject. These include 
[\refcite{Emparan:2006mm,Emparan:2008eg,Obers:2008pj}] and 
[\refcite{Kol:2004ww,Harmark:2005pp,Harmark:2007md}]. Our goal here will 
be a more modest one -- to provide an illustration of some of the key lessons 
in this field, outline some of the most interesting future directions and equip 
the reader with a first guide to the literature.

We will focus on stationary black holes solutions in pure Einstein 
gravity with simple Kaluza-Klein or flat Minkowski space asymptotics. This is a 
minimal setup, which one can envision extending in many different directions, 
$e.g.$ change the asymptotics, add more fields, more charges, supersymmetry $etc.$
Despite the apparent simplicity of our setup (indeed solutions in this setup
will be classified by a small number of asymptotic charges), we
will discover that by changing the spacetime dimension a rich pattern of 
interconnecting phases with different horizon topology and stability properties
arises.

\subsection{Why study gravity in higher dimensions?}

Classical general relativity in four spacetime dimensions has an obvious
motivation: we live in an observably four-dimensional world. Why then 
venture into a different dimension? There are several good reasons for that:
\begin{itemize}
\item[$(i)$] String theory contains gravity and typically requires more than 
four dimensions. In fact, spacetime dimension itself is a dynamical concept in 
string theory. String theory may not have proven itself as the true theory of
the nature so far, but it has had considerable success in explaining the
microscopic degrees of freedom of black holes. The first successful statistical
counting of black hole entropy was performed for a five-dimensional
black hole [\refcite{Strominger:1996sh}].
\item[$(ii)$] The AdS/CFT correspondence relates a gravitational theory
on asymptotically AdS spaces in $d+1$ dimensions to a non-gravitational
quantum field theory (QFT) in $d$ dimensions. Black holes in this context 
are useful for the analysis of the finite temperature properties of the QFT.
\item[$(iii)$] If large extra dimensions 
[\refcite{Arkani-Hamed:1998rs,Antoniadis:1998ig}] are realized in nature, higher
dimensional black holes may arise as important physical objects with 
observable signatures in accelerators or in the universe (for a review see 
[\refcite{Kanti:2004nr}]).
\end{itemize}

Besides any potential applications classical general relativity in higher
dimensions has an interest of its own. As we change the number of 
spacetime dimensions we discover new features with no counterpart
in lower dimensions. The appearance of some differences is perhaps
expected. After all, by increasing the number of spacetime dimensions we 
increase the number of degrees of freedom of the graviton and it is natural
to expect that the theory will become more versatile. What is interesting to 
distinguish are precisely those properties of the theory that are dimension-dependent.

\subsection{Novel features}

In the following sections, we will encounter various instances where
the standard lore of pure gravity in four dimensions breaks down as
the spacetime dimension increases. Two of the most salient new features
that we will encounter are the following.

\subsubsection{Black hole (non-)uniqueness and horizon topology}

The solutions of the vacuum Einstein equations $R_{\mu\nu}=0$ are
constrained by powerful uniqueness theorems in four dimensions.
A black hole is uniquely specified by the ADM mass $M$ and the
angular momentum $J$ measured at infinity 
[\refcite{Israel:1967wq,Carter:1971,Hawking:1972vc,Robinson:1975}]. 
The unique stationary solution is the Kerr black hole
\begin{subequations}
\label{novelaaA}
\begin{equation}
\label{novelaa}
ds^2=-dt^2+\frac{\mu r}{\Sigma}(dt+a \sin^2\theta d\phi)^2
+\frac{\Sigma}{\Delta}dr^2+\Sigma d\theta^2+(r^2+a^2)\sin^2\theta d\phi^2
~,
\end{equation}
\begin{equation}
\Sigma=r^2+a^2\cos^2\theta~, ~ ~ \Delta=r^2-\mu r+a^2~, ~~
\mu=2GM~, ~ ~ a=\frac{J}{M}
~.
\end{equation}
\end{subequations}
$G$ is Newton's constant. For $J=0$ we recover the unique static black
hole solution: the Schwarzschild black hole.

The generalization of the Kerr black hole solution in more than four dimensions
is provided by the Myers-Perry solution [\refcite{Myers:1986un}]. For a certain 
range of mass and angular momentum this is not anymore the unique stationary 
solution. The first explicit example of this violation of a uniqueness theorem was found
by Emparan and Reall [\refcite{Emparan:2001wn}] in five dimensions. The new 
solution that does not exist in four dimensions is a black ring, $i.e.$ a black hole 
with horizon topology $S^2 \times S^1$. 

The black ring solution brings forth the violation of another uniqueness
theorem that holds in four dimensions: the horizon topology uniqueness
theorem. In four dimensions (with flat space asymptotics), one can show 
on general grounds that the horizon topology is unique. Hawking's 
horizon topology theorem [\refcite{Hawking:1972vc,Hawking:1973uf}] 
states that the integrated Ricci scalar over the two-dimensional horizon 
should be a positive number. Hence, the horizon topology can
only be a sphere. This is no longer the case in more than four dimensions.
Generalizations of Hawking's horizon topology theorem have been considered
in [\refcite{Helfgott:2005jn,Galloway:2005mf}]. The resulting restrictions 
on topology turn out to be less restrictive
in higher dimensions. For example, one can show that the horizon 
must be positive Yamabe type, $i.e.$ it must be a manifold with an 
everywhere positive Ricci scalar. In five dimensions the only allowed
horizon topologies seem to be $S^3$ and $S^2\times S^1$. In six
dimensions many more choices appear to be allowed by topology, 
$e.g.$ $S^4$, $S^3 \times S^1$, or even $S^2\times \Sigma_g$, where
$\Sigma_g$ is a two-dimensional Riemann surface with genus $g$. 
In more than six dimensions even a partial classification of allowed
horizon topologies is currently lacking.

\subsubsection{Stability and transitions}

Another intriguing feature of higher dimensional gravity is the
ubiquitous appearance of horizon instabilities and the dynamical
transition between phases with different entropy and horizon topology.
An example that arises in various contexts is the Gregory-Laflamme (GL)
instability [\refcite{Gregory:1993vy,Gregory:1994bj}], a long-wavelength 
instability that arises when the horizon has extended directions. In that 
case, the unstable mode evolves towards breaking up the horizon into 
smaller pieces. The possibility of the formation of a naked singularity 
during this process points towards a violation of the Cosmic Censorship 
Hypothesis. A related feature is the presence of merger points. These 
are points in the black hole phase diagrams, where black holes with different 
horizon topology meet in a topology changing transition.

Another interesting property of higher dimensional gravity is the appearance,
in some cases, of a critical spacetime dimension, where some properties of 
the above patterns may change abruptly as we change the dimension. 

Finally, various bounds that exist in four dimensions (like the Kerr 
bound $J\leq GM^2$) are lifted in higher dimensions. It is possible,
although not exhibited in full generality, that a dynamical version of these 
bounds and a new version of black hole uniqueness is reinstated in 
higher dimensions by demanding stability [\refcite{Kol:2002dr}]. The 
issue of perturbative stability becomes more complicated when additional 
fields, besides the graviton, are present.

\subsection{Solution generating techniques}
\label{solutiongenerating}

The reasons that are responsible for the richness of higher dimensional
gravity are also responsible for its complexity and our limited efficiency in 
uncovering its full breadth of solutions. Here are some of the methods that 
have been used successfully in this endeavor.

\subsubsection{Exact methods}

The complexity of the equations can be reduced, in general, by assuming 
certain symmetries. In such cases, a clever ansatz for the form of the solution,
or a clever choice of the coordinate system may simplify the equations
considerably. An example of this strategy is the Weyl ansatz 
[\refcite{Emparan:2001wk,Harmark:2004rm}] for static and stationary solutions 
with $D-2$ commuting Killing vectors in $D$ dimensions. This method
has been used successfully to find exact solutions in five and six-dimensional
Kaluza-Klein (KK) spaces, and rotating black ring solutions in five dimensional 
asymptotically flat space.

In some cases, when an exact solution is known, many more can be generated
with methods such as the inverse scattering method
[\refcite{Belinsky:1971nt,Belinsky:1979,Belinski:2001ph,Pomeransky:2005sj}]. 
This method has been extremely efficient for five-dimensional stationary black hole 
solutions in flat space. Unfortunately, this method is not as helpful for similar purposes 
in higher dimensions. For a further solution generating mechanism see
[\refcite{Giusto:2007fx}].

\subsubsection{Approximate methods}

In certain cases, we have no analytic tools to determine the exact solution,
but we happen to know the solution is a certain regime of parameters. Then,
one may attempt to find the solution with a perturbative expansion around
the known limit. An example of such a method is the matched asymptotic 
expansion that has been employed succesfully in 
[\refcite{Harmark:2003yz,Gorbonos:2004uc,Karasik:2004ds},
\refcite{Gorbonos:2005px,Dias:2007hg,Emparan:2007wm}]. This method 
applies in problems that contain two (or more) widely separate scales. 
We will review its basic features in section \ref{phases}.

\subsubsection{Numerical work}

When the analytical methods fail, the only way to proceed is by numerical
methods. In the context of KK black holes these techniques have been successfully 
applied in [\refcite{Gubser:2001ac,Wiseman:2002zc,Sorkin:2004qq,Kleihaus:2006ee}
\refcite{Sorkin:2006wp,Kleihaus:2007cf}] for non-uniform black strings and 
in [\refcite{Sorkin:2003ka,Kudoh:2003ki,Kudoh:2004hs}] for localized black holes
(see section \ref{KK} for the definition of these objects). The combination
of approximate (semi-exact) analytical methods and numerical analysis
may be our best bet in generic situations where exact solutions seem out
of reach.

\subsection{A short outline}

In this review we will focus on black hole solutions of the vacuum Einstein 
equations $R_{\mu\nu}=0$ in $D\geq 4$ dimensions. Section \ref{KK} reviews 
the case of KK asymptotics. Section \ref{phases} reviews the case of stationary 
rotating black hole and black ring solutions in flat space with a single angular 
momentum. Several interesting open problems are summarized in the final 
section \ref{final}.

\section{Kaluza-Klein black holes}
\label{KK}

Kaluza-Klein black holes will provide our first illustration of the novel 
features of higher dimensional gravity. Our definition of Kaluza-Klein 
black holes in $d+p$ dimensions will be the following: vacuum solutions of the pure gravity 
Einstein equations with at least one event horizon that asymptote at infinity to 
$d$-dimensional Minkowski spacetime times a $p$-torus $({\cal M}^d \times 
{\mathbb T}^p)$. For concreteness, we will mostly discuss here static and neutral 
solutions in the $p=1$ ($i.e.$ the $S^1$) case. 

We will discover that certain aspects of the KK discussion in this section 
are instrumental in uncovering some of the crucial properties 
of rotating black holes and black rings in flat space in the next section.

A lengthy review of complementary aspects of the subject presented in this section 
can be found in [\refcite{Kol:2004ww,Harmark:2005pp,Harmark:2007md}].

\subsection{The parameters}

For any spacetime which asymptotes to ${\cal M}^d \times S^1$ we can 
define the mass $M$ and tension ${\cal T}$. These two asymptotic
quantities can be used to parametrize the various phases of KK
black holes in a $(\mu,n)$ phase diagram, which will be defined
in a moment.

For any localized static object the mass and the tension can be 
determined from the asymptotics of the metric
\begin{equation}
\label{paramaa}
g_{tt}\simeq -1+\frac{c_t}{r^{d-3}}~, ~ ~ 
g_{zz}=\simeq 1+\frac{c_z}{r^{d-3}}
\end{equation}
as $r\to \infty$. We are parametrizing the asymptotic spacetime 
${\cal M}^d\times S^1$ by the variables $t,x^1,...,x^{d-1}$ and $z\sim z+L$.
The radial variable $r=\sqrt{\sum_i (x^i)^2}$. In this notation, the mass $M$ 
and tension ${\cal T}$ are given by [\refcite{Harmark:2003dg,Kol:2003if}]
\begin{equation}
\label{paramab}
M=\frac{\Omega_{d-2}L}{16\pi G}\left[ (d-2)c_t-c_z\right]~, ~ ~ 
{\cal T}=\frac{\Omega_{d-2}}{16\pi G}\left[ c_t-(d-2)c_z\right]
\end{equation}
where $\Omega_d=\frac{2\pi^{\frac{d+1}{2}}}{\Gamma(\frac{d+1}{2})}$ 
is the volume of a unit $d$-sphere. For more details on the gravitational
tension of black holes and branes we refer the reader to 
[\refcite{Harmark:2004ch,Harmark:2004ws,Myers:1999ps,Traschen:2001pb}
\refcite{Townsend:2001rg,Kastor:2006ti}].

It will be convenient to define the following set of dimensionless quantities
\begin{equation}
\label{paramac}
\mu=\frac{16\pi G}{L^{d-2}}M~, ~ ~ 
\textsf{s}=\frac{16\pi G}{L^{d-1}}S~, ~ ~
n=\frac{{\cal T}L}{M}
~,
\end{equation}
where $S$ is the entropy. For later purposes it will also be useful to define 
the quantities
\begin{equation}
\label{paramad}
\ell=\mu^{-\frac{1}{D-3}}~, ~ ~ a_H=\mu^{-\frac{D-2}{D-3}}\textsf{s}
\end{equation}
where $D$ is the total spacetime dimension (here $D=d+1$).

The program set forth in [\refcite{Harmark:2003dg,Harmark:2003eg}] is to 
plot all phases of Kaluza-Klein black holes in a $(\mu,n)$ diagram.
One can show on general grounds [\refcite{Harmark:2003dg}] that the 
relative tension $n$ should always lie within the range $0\leq n\leq d-2$.
The upper bound follows from the Strong Energy Condition. The lower
bound was found in [\refcite{Traschen:2003jm,Shiromizu:2003gc}].

\subsection{Black holes and strings on ${\cal M}^d \times S^1$}

\begin{figure}[t]
\centerline{\psfig{file=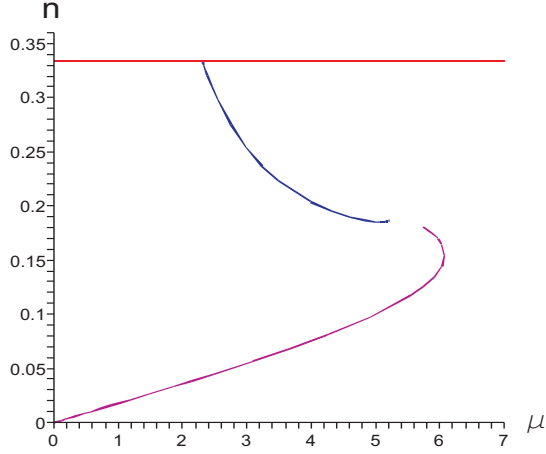,height=6.5cm,width=7.5cm}}
\vspace*{8pt}
\caption{Black hole and string phases for $d=5$ in a $(\mu,n)$ phase
diagram. The horizontal (red) line at $n=1/(d-2)=1/3$ is the uniform string 
branch. The (blue) branch emanating at the Gregory-Laflamme point is
the non-uniform string branch. It was obtained numerically in [37].
The (purple) branch starting at the point $(\mu,n)=(0,0)$ is the localized 
black hole branch. It was obtained numerically in [44].
The numerical data suggest that the non-uniform and localized branches 
meet at a horizon topology changing merger point. Figure reprinted from
Ref.~[6].
\label{figKK}}
\end{figure}

There are three main types of neutral and static black objects on 
${\cal M}^d\times S^1$ with a single connected horizon and 
$SO(d-1)$ symmetry -- the uniform black string, the non-uniform 
black string and the localized black holes. The former two have an 
event horizon with topology $S^{d-2}\times S^1$ and the latter 
horizon topology $S^{d-1}$. All these phases are depicted for $d=5$
in Fig.~\ref{figKK}, which is based on the numerical results of 
Refs.~[\refcite{Wiseman:2002zc,Kudoh:2004hs}]. The most prominent
features of this diagram are the following.

\vspace{.3cm}
\noindent{\bf The uniform black string and Gregory-Laflamme instabilities.} 
The metric of the uniform black string solution is 
\begin{equation}
\label{KKaa}
ds^2=-f dt^2+f^{-1}dr^2+r^2 d\Omega_{d-2}^2+dz^2~, ~ ~ 
f=1-\frac{r_0^{d-3}}{r^{d-3}}
~,
\end{equation}
where $d\Omega_{d-2}^2$ is the metric element of a $(d-2)$-dimensional
unit sphere. This metric is simply the direct sum of the $d$-dimensional
Schwarzschild-Tangherlini static black hole with an extra flat $z$ direction.
This phase has constant relative tension $n=\frac{1}{d-2}$.

A crucial feature of the uniform black string solution is the Gregory-Laflamme 
instability -- a long-wavelength instability [\refcite{Gregory:1993vy,Gregory:1994bj}]
associated to a mode that propagates along the direction of the string (the $z$ direction)
and grows exponentially with time. The instability occurs for any $\mu<\mu_{\rm GL}$.
For $\mu>\mu_{\rm GL}$ the string is believed to be classically stable. For 
$\mu=\mu_{\rm GL}$ there is a marginal, static mode which signals the emergence
of a new branch of black string solutions that are non-uniform along the circle.
An analytic formula for the Gregory-Laflamme mass $\mu_{\rm GL}$ is not
known. Numerical estimates can be found in 
[\refcite{Gregory:1993vy,Gubser:2001ac,Sorkin:2004qq}] and are summarized in 
[\refcite{Harmark:2007md}]. A large $d$ analytical expression was derived in 
[\refcite{Kol:2004pn}].

\vspace{.3cm}
\noindent{\bf The non-uniform black string, the localized black hole and a merger.}
The new non-uniform branch emanating at $\mu_{\rm GL}$ (the blue line in 
Fig.~\ref{figKK}) has the same horizon topology as the uniform black string 
$S^{d-2}\times S^1$, but a smaller relative tension $n<\frac{1}{d-2}$. 
The analytic form of the non-uniform black string is not known. Perturbatively 
around $\mu_{\rm GL}$ one finds that the relative tension behaves as
\begin{equation}
\label{KKab}
n(\mu)=\frac{1}{d-2}-\gamma(\mu-\mu_{\rm GL})+
{\cal O}\left((\mu-\mu_{\rm GL})^2\right)
~.
\end{equation}
The qualitative behavior of the non-uniform branch depends on the sign of 
$\gamma$. Surprisingly, it turns out [\refcite{Sorkin:2004qq}] that there is a 
critical dimension $D=d+1=14$ where $\gamma$ changes sign. $\gamma$
is positive for $d\leq 12$ and negative for $d\geq 13$.

The phase diagram in Fig.~\ref{figKK} includes an additional branch (the purple 
branch) that comes off the origin $(\mu,n)=(0,0)$. This is the branch of a black hole 
localized in the $S^1$. It has horizon topology $S^{d-1}$. The existence of such 
solutions is intuitively clear. In the small mass limit $\mu\ll 1$ and to leading order, 
this is simply a $(d+1)$-dimensional Schwarzschild black hole. The presence of the 
periodic boundary conditions in $z$ deform the finite $\mu$ solution away from 
Schwarzschild. The exact analytic form of this branch is also unknown. It is possible, 
however, to analyze the metric of small black holes analytically by using the method 
of asymptotic expansions [\refcite{Harmark:2003yz,Gorbonos:2004uc,Gorbonos:2005px}]. 
This method sets up a perturbative expansion around the small mass limit $\mu \ll 1$. 
A similar approach in the next section will teach us about higher dimensional black 
rings in an ultra-spinning regime.

Fig.~\ref{figKK} suggests that the above two branches meet at a topology changing
transition point. The scenario of such a transition was suggested early on by Kol 
[\refcite{Kol:2002xz}]. More details about the physics of the merger point can be 
found in [\refcite{Kol:2002xz,Kol:2004ww}].

\vspace{.3cm}
\noindent{\bf Other phases.} There are several other phases that supplement the 
$(\mu,n)$ phase diagram in Fig.~\ref{figKK}. These include:

\vspace{.2cm}
\noindent
{$\bullet$ \bf Copies.} In [\refcite{Harmark:2003eg}] it was argued that one can 
generate new solutions by copying solutions on the circle several times following 
an idea in [\refcite{Horowitz:2002dc}]. For any positive integer $k$ the new solution 
has mass and relative tension $\tilde \mu=\frac{\mu}{k^{d-3}}$, $\tilde n=n$. More 
details can be found in [\refcite{Harmark:2007md}] and the original references therein.

\vspace{.2cm}
\noindent
{$\bullet$ \bf Kaluza-Klein bubbles.} All solutions in Fig.~\ref{figKK} lie within the range
$0\leq n \leq \frac{1}{d-2}$. Solutions in the range $\frac{1}{d-2} <n \leq d-2$ were
argued to exist in [\refcite{Elvang:2004iz}] (see also [\refcite{Emparan:2001wk,Elvang:2002br}]). 
They comprise of the static Kaluza-Klein bubble and bubble-black hole sequences. 
Further details on these solutions can be found in [\refcite{Harmark:2007md}] and 
the original references therein.

\vspace{.2cm}
\noindent
{$\bullet$ \bf Multi-black hole solutions.} Multi-black hole solutions were considered
recently in Ref.~[\refcite{Dias:2007hg}] using the method of asymptotic expansions.

\begin{figure}[t!]
\centerline{\psfig{file=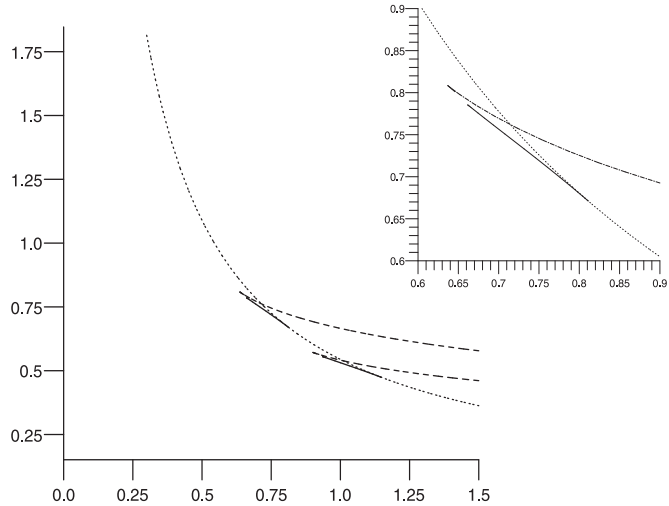,height=8cm,width=9.5cm}}
\vspace*{8pt}
\caption{A plot of the rescaled entropy $a_H$ versus rescaled mass for 
Kaluza-Klein black holes with one uniform direction in ${\cal M}^5 \times 
{\mathbb T}^2$. The dotted line represents the uniform black membrane 
phase, the solid line the non-uniform black membrane phase and the dashed 
line the localized black string phase. For the latter two phases we also depict the 
$k=2$ copy. The non-uniform black membrane phase emanates from the 
uniform phase at the Gregory-Laflamme point $\ell_{\rm GL}=0.811$, while the 
$k=2$ copy starts at $\ell^{(2)}_{\rm GL}=\sqrt 2 \ell_{\rm GL}=1.15$. A zoom of 
the non-uniform phase region appears in the diagram in the right upper corner.
This figure is representative of the ${\cal M}^{D-2}\times {\mathbb T}^2$
for $6\leq D\leq 14$. Figure reprinted from Ref.~[35].
\label{figKK7d}}
\end{figure}

\subsection{A note on black membranes and black strings on ${\cal M}^d \times {\mathbb T}^2$}
\label{blacktorus}

There is a similar pattern of phases in other Kaluza-Klein spaces, $e.g.$ 
${\cal M}^d \times {\mathbb T}^p$. Clearly new possibilities arise as 
we increase the dimension of the compact manifold. An example that will 
be relevant for the discussion of the next section is the case of the torus 
($p=2$). Black membrane phases that are non-uniform in one of the directions
of the two-torus can be imported straightforwardly from the simpler
$p=1$ case that we described above. Such phases were considered in 
[\refcite{Emparan:2007wm}]. In Fig.~\ref{figKK7d}, which is taken from 
Ref.~[\refcite{Emparan:2007wm}], we plot the rescaled entropy $a_H$
in terms of the rescaled mass parameter $\ell$ (see the definition 
\eqref{paramad}) for the uniform black membrane phase (dotted line), 
the non-uniform black membrane phase (solid line), the localized black 
string phase (dashed line) and the $k=2$ copies of the latter two.

\section{Phases of neutral rotating black holes and rings in flat space}
\label{phases}

In this section we will summarize what is known about the phases of 
neutral rotating black holes in the $D$-dimensional Mikowski space 
${\cal M}^D$. Again we are considering pure gravity, so we are solving 
the vacuum Einstein equations $R_{\mu\nu}=0$. In $D$ dimensions a 
black hole can rotate in $[\frac{D-1}{2}]$ independent directions. For 
simplicity, we will consider the case of black holes with a single 
non-vanishing angular momentum, $i.e.$ black holes rotating
on a single plane. Sometimes, it will be convenient to denote the total 
dimension as $D=n+4$.

In order to compare the properties of different phases a common scale
needs to be introduced. Classical general relativity with Minkowski
space asymptotics does not possess an intrinsic scale, hence we will
take our reference scale to be one of the physical parameters of our 
solutions. We choose to take the mass and define the dimensionless 
quantities
\begin{equation}
j^{n+1} \propto \frac{J^{n+1}}{GM^{n+2}}~, ~ ~ 
a_H^{n+1} \propto \frac{{\cal A}^{n+1}}{(GM)^{n+2}}
\end{equation}
where $J$ is the angular momentum and $\cal A$ the horizon area.
Convenient proportionality constants can be found in [\refcite{Emparan:2007wm}].
We will plot different phases in a ``phase diagram'' of the rescaled
horizon area $a_H$ versus rescaled angular momentum $j$.

To show clearly how dimension affects the qualitative features of rotating 
black holes we will consider in turn the phase diagrams in four, five and
higher dimensions. The material in this section is heavily based on 
Ref.~[\refcite{Emparan:2007wm}] where the reader can find a more
detailed discussion.

\begin{figure}[t!]
\begin{picture}(0,0)(0,120)
\put(290,0){$j$}
\put(60,120){$a_H$}
\put(178,-5){$j_*$}
\end{picture}
\centerline{\psfig{file=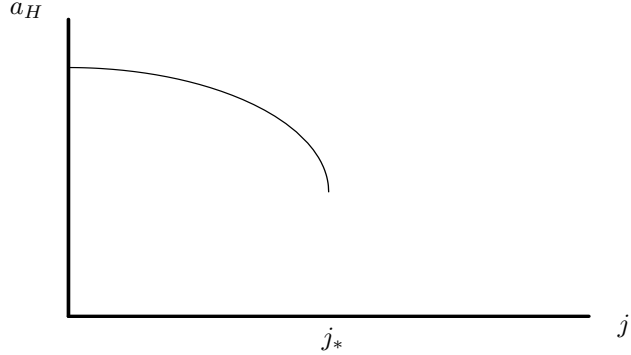,height=4cm,width=7cm}}
\vspace*{8pt}
\caption{A plot of the rescaled entropy $a_H$ versus rescaled 
angular momentum $j$ in four dimensions for rotating black hole
solutions. There is a single branch -- the Kerr black hole -- ending 
at the Kerr bound value of the angular momentum $j_*$.
\label{figKerr}}
\end{figure}

\subsection{The ``simplicity'' of four dimensions}
\label{4d}

As we reviewed in the introduction, vacuum black hole solutions 
of pure gravity are severely constrained by uniqueness theorems. 
This is amply visible in the phase diagram of Fig.~\ref{figKerr} -- 
there is a single branch, that of the Kerr black hole. The well known 
analytic form of the Kerr black hole metric \eqref{novelaaA} allows the 
explicit computation of the horizon area 
\begin{equation}
{\cal A}=8\pi G\left(GM^2+\sqrt{G^2M^4-J^2}\right)
~.
\end{equation}
The branch terminates at a maximum value of the angular momentum 
$J_*=GM^2$, the Kerr bound. At $J_*$ the black hole has finite
horizon area.

Another feature of four dimensions is the absence of stationary 
multi-black hole solutions. We will soon see that this is another property
unique to four dimensions.

\subsection{The ``finesse'' of five dimensions}
\label{5d}

\begin{figure}[t!]
\begin{picture}(0,0)(0,120)
\put(290,0){$j^2$}
\put(60,125){$a_H$}
\put(202,-5){$j_*^2$}
\put(140,90){\it MP black hole}
\put(270,35){\it black ring}
\put(270,15){\it black saturn}
\put(45, 110){$a_{H,max}$}
\end{picture}
\makebox{\centerline{\psfig{file=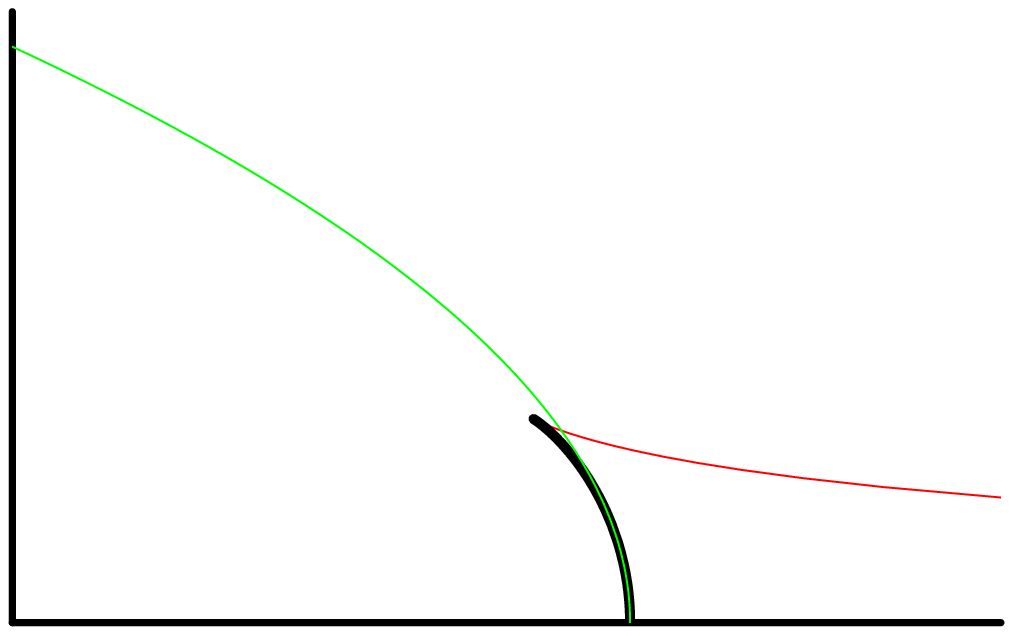,height=4cm,width=7cm}
\hspace{-7.25cm}
\psfig{file=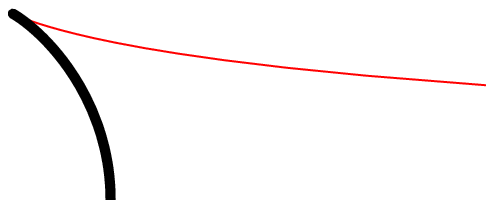,height=4cm,width=7cm}}}
\vspace*{8pt}
\caption{A plot of the rescaled entropy $a_H$ versus rescaled 
angular momentum squared $j^2$ in five dimensions for rotating 
black hole solutions. Three branches are plotted: the Myers-Perry
black hole branch (green), the black ring branch (more entropic
black plus red lines) and the black saturn branch (less entropic
black plus red lines). For the black ring/saturn the black line represents
the fat black ring/saturn and the red line the thin black ring/saturn.
It is possible that these phases exhaust the allowed phases of rotating
black holes with a single angular momentum in thermal equilibrium. 
\label{fig5D}}
\end{figure}

The phase diagram of five-dimensional neutral black holes with a 
single angular momentum is depicted in Fig.~\ref{fig5D}. It is possible
[\refcite{Elvang:2007hg}] that the phases depicted in Fig.~\ref{fig5D} 
exhaust the allowed phases of rotating black holes with a single
angular momentum in thermal equilibrium (see, however, [\refcite{Evslin:2008py}]).
Work related to the classification of the five-dimensional phases can
be found in [\refcite{Morisawa:2004tc,Hollands:2007aj}]. The main 
properties of the phases in Fig.~\ref{fig5D} are the following.

The green line extending from $J=0$ (the Schwarzschild-Tangerlini black
hole) to $J=J_*=\sqrt{\frac{32G}{27\pi}M^3}$ (a five-dimensional version 
of the Kerr bound) represents the Myers-Perry (MP) branch. This branch is the 
generalization of the Kerr black hole in five dimensions. Hence, it represents
a rotating black hole with horizon topology $S^3$. The metric is known 
analytically for the Myers-Perry black hole [\refcite{Myers:1986un}]. The 
presence of a maximal angular momentum is common in four and five 
dimensions, however, in five dimensions the maximally spinning black 
hole has zero horizon area, $i.e.$ it exhibits a naked singularity.

Besides the MP branch, Fig.~\ref{fig5D} includes two more phases:
the black ring [\refcite{Emparan:2001wn}] and the black saturn 
phases [\refcite{Elvang:2007rd}]. Both of them have two
components -- one depicted by a thick black line (commonly known
as the fat black ring/saturn), and another depicted by a thin red line
that goes off to infinite angular momentum (the latter is known as the
thin black ring/saturn). The black ring branch is the most entropic 
phase at large angular momentum. It represents a rotating black hole 
solution with a single connected horizon of topology $S^2\times S^1$. 
The black saturn branch has a disconnected horizon with two 
components: a central horizon with topology $S^3$ and an outer horizon 
with topology $S^2\times S^1$. This is an object in static equilibrium 
that comprises of a central rotating MP black hole and an outer rotating 
black ring. The two disconnected horizons are in thermodynamic 
equilibrium, $i.e.$ they have the same temperature and horizon 
angular velocity.

The neutral black rings and saturns that we consider here have several
types of instabilities. A complete classical stability analysis is complicated and
has not been performed for these objects, however, in certain regimes it 
is known that these objects exhibit instabilities. For example, an ultra-spinning
thin black ring is well-approximated by a boosted black string and boosted
black strings are known to suffer from Gregory-Laflamme instabilities
[\refcite{Emparan:2001wn,Hovdebo:2006jy}]. A fat black ring is also expected to suffer 
from instabilities under radial perturbations [\refcite{Elvang:2006dd}]. 
It is unclear whether there is an intermediate regime of angular momentum 
where the black ring is stable. Similarly, black saturns are expected to suffer 
from instabilities. Some of them are instabilities of the outer black ring, others 
are instabilities special to the black saturn. For instance, moving the central 
MP black hole off the center is expected to destabilize the saturn.

More phases can be added to the phase diagram in Fig.~\ref{fig5D} as soon as
we relinquish the assumption of thermodynamic equilibrium. Then, one
can argue for the existence of continuous families of multi-black hole solutions
that cover any point in the strip with $0\leq j<\infty$ and $0<a_H<a_{H,max}$
($a_{H,max}$ is the area of a Schwarzschild black hole)
[\refcite{Elvang:2007hg}]. At any non-zero $j$, the maximum area is achieved 
by black Saturns having an almost static central black hole that carries most of 
the mass, and a very thin and long black ring that carries most of the angular 
momentum.

The presence of all these new phases demonstrates clearly the lessons
we outlined in the introduction. The uniqueness theorems that were so
powerful in four dimensions are now explicitly violated and the new phases
exhibit novel properties, $e.g.$ new horizon topologies, stationary multi-black
hole configurations $etc$.

\subsection{The abyss of higher dimensions}
\label{higherd}

Five dimensions are special for several reasons. First of all,
in five dimensions the complexity of the Einstein equations is still under 
considerable control. Symmetries and the solution generating techniques 
of section \ref{solutiongenerating} allow for a fairly complete picture of the 
allowed phases. This is not the case for six dimensions and higher, where 
semi-quantitative methods are currently the major source of information. 
The still illusive discovery of the exact metric of a higher dimensional black 
ring is a characteristic example of the increasing complexity of higher 
dimensional gravity. A nice general discussion on the difficulties of higher 
dimensional gravity can be found in the introductory section of 
Ref.~[\refcite{Emparan:2008eg}].

The special nature of five dimensions is qualitatively apparent also in 
some of the key properties of the Myers-Perry solutions, which are known
analytically in any dimension $D$. What controls the key qualitative properties 
of the MP solution is a competition between the centrifugal repulsion 
$\frac{J^2}{M^2r^2}$, which is dimension independent, and the gravitational 
attraction $-\frac{GM}{r^{D-3}}$, which is dimension dependent. For $D=4$
the gravitational attraction is the contribution that dominates at large distances. 
For $D>5$ it is the centrifugal repulsion. For $D=5$ the two contributions are 
comparable. 

\begin{figure}[t!]
\begin{picture}(0,0)(0,120)
\put(340,-50){$j$}
\put(20,120){$a_H$}
\put(60,-55){$j_{mem}$}
\put(70,90){\it MP black hole}
\put(270,25){\it thin black ring}
\end{picture}
\centerline{\psfig{file=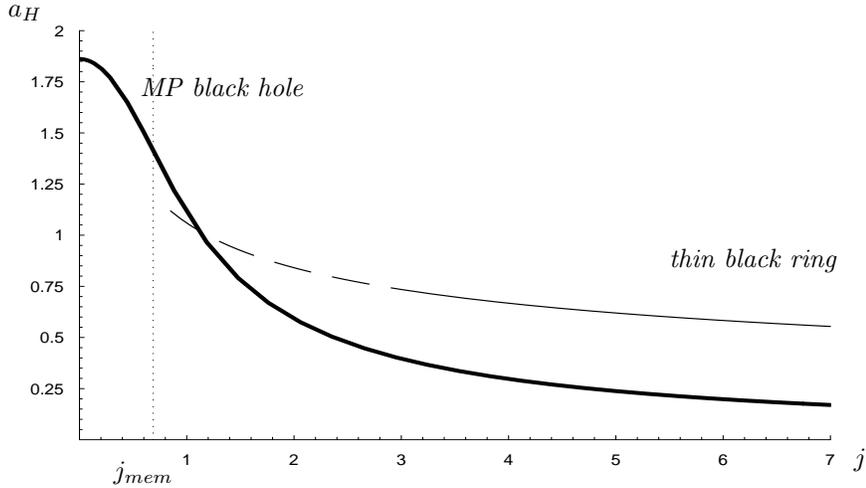,height=6cm,width=11cm}}
\vspace*{8pt}
\caption{Area vs spin for fixed mass, $a_H(j)$, in seven dimensions. The
thick curve is the exact result for the MP black hole. The vertical dotted line
intersects this curve at the inflection point $j_{mem}=2^{1/4}/\sqrt 3$, $a_H=\sqrt 2$.
It signals the onset at larger $j$ of membrane-like behavior for MP black holes. The
thin curve represents thin black rings which have been analyzed via the method
of asymptotic expansions at large $j$. Figure reprinted from Ref.~[35].
\label{figMPBR}}
\end{figure}

The Myers-Perry curve is depicted as a solid black line in Fig.~\ref{figMPBR}
(for $D=7$). The most notable new feature of the $D>5$ MP curves is the absence of
a Kerr bound. In fact, the curve changes its Kerr-like character at large enough
angular momentum. The point where this change takes place can be approximated
by the inflection point $j_{mem}$ where $\partial^2 a_H/\partial j^2=0$. It has
been argued [\refcite{Emparan:2003sy}] that the ultra-spinning regime of the
Myers-Perry black hole is captured by a $static$ black membrane solution. 
The horizon spreads out along the plane of rotation with a size proportional
to the angular momentum. We will return to this point in a moment.

\subsection{Forging the black ring}
\label{blackring}

Besides the Myers-Perry black holes it is natural to expect on the
basis of the five-dimensional example that black rings are also 
part of the $D$-dimensional phase diagram. In the absence of 
exact black ring solutions in arbitrary dimensions, 
Ref.~[\refcite{Emparan:2007wm}] proceeded to construct approximate
black ring solutions around an ultra-spinning regime using the method
of asymptotic expansions.

Black rings in any dimension $D$ are objects of horizon topology 
$S^{D-3}\times S^1$. They are characterized by two dimensionful
parameters: $r_0$ (roughly the fatness of the horizon) and $R$
(the $S^1$ radius of the horizon). The physical intuition behind a 
black ring is the simple intuitive picture that a black ring arises 
from a straight black string by bending the string and spinning it to 
balance the circular ring. This picture becomes more precise in the 
ultra-spinning regime.

In the regime of large angular momentum the ring becomes long and 
thin and the two scales $r_0, R$ become widely separated, $i.e.$ $r_0/R\ll 1$. 
In this case, a black ring is well approximated by a boosted black string. This is 
an object with horizon topology $S^{D-3}\times {\mathbb R}$ and metric
(recall $n=D-4\geq 1$)
\begin{eqnarray}
\label{boosted}
ds^2&=&-\left( 1-\cosh^2 \alpha \frac{r_0^n}{r^n}\right)
-2\frac{r_0^n}{r^n} \cosh\alpha \sinh\alpha dtdz
\nonumber\\
&&+\left(1+\sinh^2\alpha \frac{r_0^n}{r^n}\right)dz^2+
\left(1-\frac{r_0^n}{r^n}\right)^{-1}dr^2+r^2 d\Omega_{n+1}^2
~.
\end{eqnarray}
$\alpha$ is a boost parameter whose precise value we will 
determine in a moment.

The basic idea behind the method of matched asymptotic expansions
is to setup a perturbative scheme where we solve the equations of 
motion independently in the near-horizon and far-away regions of the black
object and then match the two expansions in the overlap region
which is large when $r_0\ll R$. The near-horizon expansion is
an expansion in powers of $r/R$ around the boosted black string solution
\eqref{boosted}, whereas in the far-away region we expand in powers of 
$r_0/r$. 

Ref.~[\refcite{Emparan:2007wm}] performed this expansion to
first order and found an explicit solution which describes, to this order,
a thin black ring solution in arbitrary dimension. In the process, one
finds that regularity of the solution forces a single value of the boost
parameter $\alpha$ in \eqref{boosted}. This is
\begin{equation}
\label{balance}
\sinh^2 \alpha=\frac{1}{n} ~ \Leftrightarrow ~ 
R=\frac{n+2}{\sqrt{n+1}}\frac{J}{M}
~.
\end{equation}
The second equality between the different parameters of the boosted
black string expresses the balancing condition of the ring. This condition
is equivalent to the vanishing of the tension of the boosted black string,
$i.e.$ $T_{zz}=0$. This is a special case of the Carter equations
[\refcite{Carter:2000wv}]
\begin{equation}
\label{carter}
K^{\rho}_{\mu\nu}T^{\mu\nu}=0
\end{equation}
which follow from the conservation law of the stress-energy momentum 
tensor. $K^\rho_{\mu\nu}$ is the second fundamental tensor, an object
that extends the notion of extrinsic curvature to submanifolds of 
co-dimension larger than one. 

The method of asymptotic expansions and the Carter equations \eqref{carter} 
are also a useful tool in the exploration of black objects with more general 
horizon topologies [\refcite{Emparan:2008hno}]. \footnote{An early report on 
the results of [\refcite{Emparan:2008hno}] can be found in 
[\refcite{Emparan:2008edi}].} Recently, these tools have been used 
successfully to extend the above analysis to black rings in (A)dS 
[\refcite{Caldarelli:2008pz}].

\subsection{Towards a complete phase diagram}
\label{phasediagram}

The above results demonstrate the existence of a thin black ring in arbitrary 
dimensions. We plotted this new phase as a thin line in Fig.~\ref{figMPBR}.
Notice that the black ring is more entropic than the MP black hole in 
the ultra-spinning regime.

\begin{figure}[t!]
\begin{picture}(0,0)(0,120)
\put(15,90){${uniform \atop {black~ membrane}}$}
\put(15,50){${non-uniform \atop {black~ membrane}}$}
\put(15,10){${horizon \atop {pinches~ off}}$}
\put(15,-40){${localized \atop {black~ string}}$}
\put(280,90){${ultra-spinning \atop {MP~ black~ hole}}$}
\put(280,50){${pinched \atop {black~ hole}}$}
\put(280,10){${horizon \atop {pinches~ off}}$}
\put(280,-35){${~ \atop {black~ ring}}$}
\end{picture}
\centerline{\psfig{file=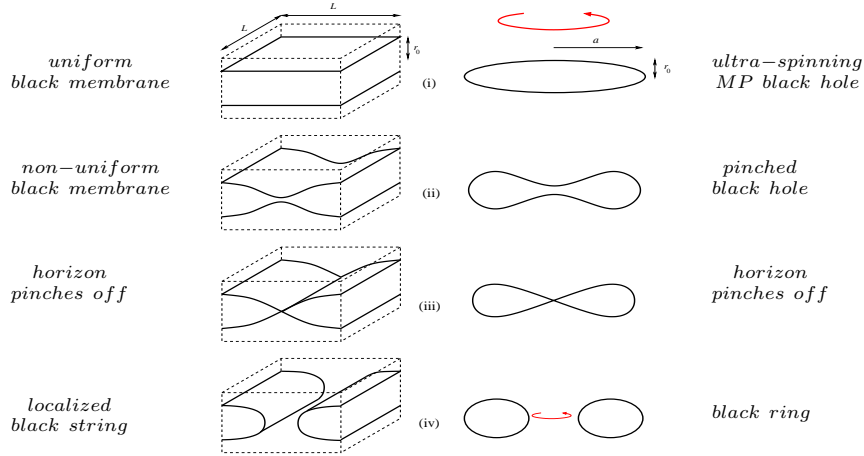,height=6cm,width=6cm}}
\vspace*{8pt}
\caption{The correspondence between phases of black membranes
wrapped on a ${\mathbb T}^2$ of side $L$ (left) and ultra-spinning MP
black holes with rotation parameter $a\sim \frac{J}{M}$ (right: must be 
rotated along a vertical axis). Reprinted from Ref.~[35].
\label{figmembranes}}
\end{figure}

\begin{figure}[t!]
\begin{picture}(0,0)(0,120)
\put(340,-50){$j$}
\put(20,120){$a_H$}
\put(60,-55){$j_{mem}$}
\put(70,90){\it MP black hole}
\put(270,25){\it thin black ring}
\end{picture}
\centerline{\psfig{file=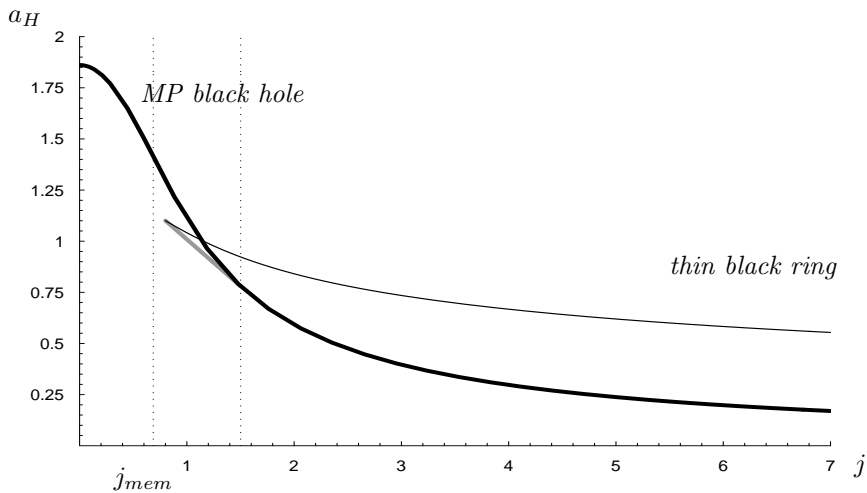,height=6cm,width=11cm}}
\vspace*{8pt}
\caption{Qualitative completion of Fig.~\ref{figMPBR} using Fig.~\ref{figKK7d}.
The gray line corresponds to the conjectured phase of pinched black
holes which branch off tangentially from the MP curve (thick) at a value
$j_{\rm GL}>j_{mem}$ and merge with the black ring curve (thin). At
any given dimension, the phases may not display the swallowtail in 
Fig.~\ref{fig7dphases}, depicted here. Reprinted from Ref.~[35].
\label{fig7dphases}}
\end{figure}

A nice picture for the rest of the phase diagram follows from the 
ensuing observation. We noticed above that the MP black hole
is approximated well beyond the critical spin $j_{mem}$ by a black
square torus with size $L\sim \frac{J}{M}$ and $S^{D-4}$ size $r_0$.
In section \ref{KK} we reviewed the basic features of the phase diagram
of black strings and branes. Hence, the analogy between ultra-spinning 
MP black holes and black membranes suggests the picture summarized
in Fig.~\ref{figmembranes}. The pinching of the horizon of a non-uniform 
black membrane to the horizon of a localized black string corresponds to the 
pinching of the horizon of an ultra-spinning MP black hole to the horizon 
of black ring. Thus, importing the black membrane phase diagram 
in Fig.~\ref{figKK7d} to the phase diagram of rotating black holes in 
$D$ dimensions suggests the completion of Fig.~\ref{figMPBR} that 
appears in Fig.~\ref{fig7dphases} (in this analogy associate the KK mass 
$\mu$ with the MP spin $j$). The analog of the Gregory-Laflamme mode
on the MP branch signals the existence of a pinched black hole which
merges with the thin black ring branch.

\begin{figure}[t!]
\begin{picture}(0,0)(0,120)
\put(340,-60){$j$}
\put(10,120){$a_H$}
\end{picture}
\centerline{\psfig{file=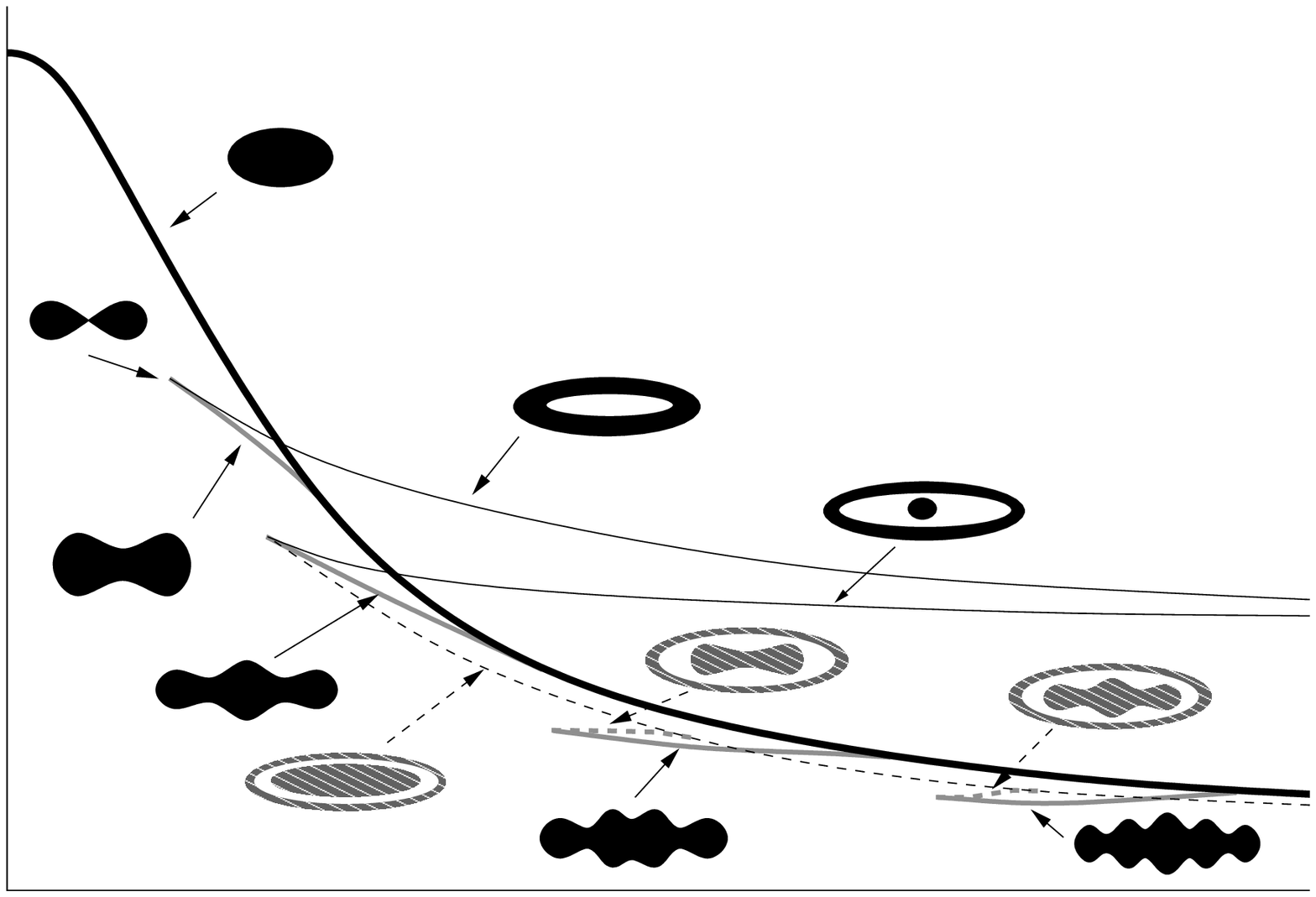,height=6cm,width=11cm}}
\vspace*{8pt}
\caption{Proposal for the phase diagram of thermal equilibrium phases in 
$D\geq 6$. The solid lines and figures have significant arguments in their
favor, while the dashed lines and figures might not exist and admit conceivable,
but more complicated, alternatives. Reprinted from Ref.~[35].
\protect\label{figfullphase}}
\end{figure}

The analogy with Fig.~\ref{figKK7d} suggests the existence of further 
GL-like points along the MP branch. From each of these points a {\it pinched 
black hole} with more and more pinches comes off. The second GL 
point gives rise to a MP black hole that develops a circular pinch, which
then grows deeper until the merger with a black Saturn configuration in 
thermal equilibrium (see Fig.~\ref{figfullphase}).

There is a possibility that a second kind of black Saturn, also
in thermal equilibrium, exists in $D\geq 6$, which would not have a 
counterpart in five dimensions. Indeed, in our phase diagrams, we fix the
total $M$ and $J$ of the black Saturn, but the mass and spin of its two
constituents are not fixed separately. Instead, under thermal equilibrium
we demand only that the temperature and angular velocity of the black ring
and the central black hole are equal.  So, besides the black Saturn with 
a small, round central black hole, that we have discussed above, it may
be possible to have another one with a large, pancaked central black
hole. With the assumption that these {\it pancaked black Saturns} exist,
Ref.~[\refcite{Emparan:2007wm}] proposed a picture where the pancaked
black Saturn branch persists to infinite angular momentum and in between
exhibits GL-like zero modes from which new branches of {\it pinched 
black Saturn} solutions emerge. A possible completion of the phase diagram
where these solutions merge with the above-mentioned pinched MP black 
holes is depicted in Fig.~\ref{figfullphase}. A more detailed discussion of
this proposal and its underlying assumptions can be found in 
Ref.~[\refcite{Emparan:2007wm}]. The more recent work [\refcite{Emparan:2008hno}]
suggests that the phase of pancaked black Saturns is unlikely and hence
that this part of the phase diagram in Fig.\ref{figfullphase} needs to be 
modified.

\section{Open problems}
\label{final}

There are many aspects of higher dimensional black holes that we did
not discuss in this short review. For a more complete discussion of 
complementary aspects of higher dimensional black holes, $e.g.$ more
examples and applications, we recommend the 
Refs.~[\refcite{Emparan:2006mm,Emparan:2008eg,Obers:2008pj},
\refcite{Kol:2004ww,Harmark:2005pp,Harmark:2007md}]. We would 
like to conclude with a brief list of interesting open problems related
to the issues covered in this review. More can be found in the 
aforementioned references.

\vspace{.2cm}\noindent
{\bf Beyond the ultra-spinning regime.} It would be interesting to find 
a direct construction of the entire black ring phase and of the other 
branches proposed in Fig.~\ref{figfullphase}. Because of the complexity 
of the problem, numerical methods appear to be the most promising approach.

\vspace{.2cm}\noindent
{\bf Classical stability.} For most of the phases described in this review
a classical stability analysis has not been performed explicitly. For some
of them it is known in a suitable regime that the black hole is unstable. 
For example, thin black rings and pancaked MP black holes in the 
ultra-spinning regime are well-approximated by black strings and membranes,
which are known to suffer from Gregory-Laflamme instabilities 
[\refcite{Emparan:2001wn,Hovdebo:2006jy,Elvang:2006dd,Emparan:2003sy}]. 
However, this instability may switch off at lower spin $j\sim {\cal O}(1)$. Other 
cases where it would be nice to know if there are unstable modes include 
the non-uniform black string and the localized black hole for KK black holes
and the rotating pinched black holes in flat space.

Many features of stability change when the system includes more fields
beside the metric. It would be interesting to find general $D$ situations where 
the stability properties of the neutral solutions in this review are improved.

\vspace{.2cm}\noindent
{\bf More fields, other asymptotics.} Semi-quantitative methods, like the
method of asymptotic expansions, are useful in uncovering the properties
of black holes in more general situations with more fields, more charges
and other asymptotics. Black rings in AdS or dS spacetimes have been
studied recently in this way in [\refcite{Caldarelli:2008pz}]. We point out that 
exact five-dimensional black ring solutions with charges and dipoles have 
been constructed in [\refcite{Elvang:2003yy,Elvang:2004rt,Emparan:2004wy}].
The existence of small supersymmetric black rings in $D\geq 5$ was argued
in [\refcite{Dabholkar:2006za}]. Neutral and charged black strings in AdS
have been discussed in [\refcite{Brihaye:2007ju,Delsate:2008kw,Brihaye:2008br,Delsate:2008iv}].

\vspace{.2cm}\noindent
{\bf More angular momenta.} The phase diagrams proposed in 
Ref.~[\refcite{Emparan:2007wm}] are referring to black holes with a
single angular momentum. One may try to extend these ideas to black
rings with horizon $S^{n+1}\times S^1$ rotating not only along the
$S^1$, but also along the $S^{n+1}$. Ultra-spinning along the extra 
directions may lead to additional pinches which presumably connect
to phases with horizon topology $S^n\times S^1 \times S^1$. In five dimensions
exact solutions of doubly rotating black rings have been considered in 
[\refcite{Pomeransky:2006bd}]. Extremal multi-spinning black rings in
diverse dimensions in pure gravity were discussed in [\refcite{Figueras:2008qh}].

\vspace{.2cm}\noindent
{\bf Blackfolds -- other horizon topologies.} The basic physical intuition that
a process of bending and spinning black strings gives rise to black rings may 
be applied to more general black $p$-branes to obtain black holes with more exotic
horizon topologies -- black objects we will generally refer to as blackfolds. 
The method of asymptotic expansions and the balancing condition \eqref{carter}
are useful constructive tools in the exploration of such objects.
Progress in this direction has been achieved recently in [\refcite{Emparan:2008hno}].

\vspace{.2cm}\noindent
{\bf Plasma balls and rings.} For black holes in AdS there is also another more 
indirect approach based on the AdS/CFT correspondence. In the original
Ref.~[\refcite{Lahiri:2007ae}] a correspondence was setup between 
stationary, axially symmetric, spinning configurations of fluid in ${\cal N}=4$
super-Yang-Mills compactified to $d=3$ on a Scherk-Schwarz circle and 
large rotating black holes and black rings in AdS$_5$. The phase diagram 
of the rotating fluid exhibits many of the qualitative features of MP black holes 
and black rings in asymptotically flat space that we discussed in this review. 
Higher dimensional generalizations of this setup provide predictions for black 
holes in compactified $AdS_{D}$ with $D>5$. Further work in this direction
has appeared in [\refcite{Bhattacharyya:2007vs,Bhattacharyya:2008jc},
\refcite{Bhattacharyya:2008ji,Bhardwaj:2008if}]. There is a fruitful interplay 
between this method and more direct approaches to black holes in AdS, 
which deserves further study.

\section*{Acknowledgments}

I would like to thank Roberto Emparan, Troels Harmark, Niels Obers and 
Maria Jose Rodriguez for an enjoyable collaboration on the work presented 
here and for many interesting and enlightening discussions. Also special 
thanks to Roberto Emparan and Niels Obers for their comments on the present manuscript.
This work is partially supported by the ANR grant, ANR-05-BLAN-0079-02, 
the RTN contracts MRTN-CT-2004-005104 and MRTN-CT-2004-503369, the 
CNRS PICS \#~2530, 3059 and 3747, and by a European Union Excellence 
Grant, MEXT-CT-2003-509661.




\providecommand{\href}[2]{#2}\begingroup\raggedright\endgroup

\end{document}